\documentclass[twocolumn,showpacs,showkeys,amsmath,amssymb]{revtex4}

\usepackage{graphicx}
\usepackage{bm} 
\usepackage{subfigure}
\usepackage[T1]{fontenc}

\begin{document}
 
\title{Temperature dependence of sound velocity and hydrodynamics of ultra-relativistic heavy-ion collisions
%\footnote{\\ Research supported by the Polish State Committee for
%Scientific Research, grant 2P03B~05925}
}

\author{Miko{\l }aj Chojnacki} 
\email{Mikolaj.Chojnacki@ifj.edu.pl}
\affiliation{The H. Niewodnicza\'nski Institute of Nuclear Physics, 
Polish Academy of Sciences, PL-31342 Krak\'ow, Poland}

\author{Wojciech Florkowski} 
\email{Wojciech.Florkowski@ifj.edu.pl}
\affiliation{Institute of Physics, \'Swi\c{e}tokrzyska Academy,
ul.~\'Swi\c{e}tokrzyska 15, PL-25406~Kielce, Poland} 
\affiliation{The H. Niewodnicza\'nski Institute of Nuclear Physics, 
Polish Academy of Sciences, PL-31342 Krak\'ow, Poland}

\date{February 7, 2007}

\begin{abstract}
The effects of different forms of the sound-velocity function $c_s(T)$ on the hydrodynamic evolution of matter created in the central region of ultra-relativistic heavy-ion collisions are studied. At high temperatures (above the critical temperature $T_c$) we use the sound velocity function obtained from the recent lattice simulations of QCD, whereas at low temperatures we use the ideal hadron gas model. At moderate temperatures different interpolations between those two results are employed. They are characterized by different values of the local maximum (at $T = 0.4 \,T_c$) and local minimum (at $T=T_c$). The extreme values are chosen in such a way that at high temperature all considered sound-velocity functions  yield the entropy density consistent with the lattice simulations of QCD. We find that the presence of a distinct minimum of the sound velocity leads to a very long ($\sim$ 20 fm/c) evolution time of the system. Since such long evolution times are not compatible with the recent estimates based on the HBT interferometry, we conclude that the hydrodynamic description becomes adequate if the QCD cross-over phase transition renders the smooth temperature variations of the sound velocity, with a possible shallow minimum at $T_c$ where the values of $c_s^2(T)$ remain well above 0.1. 
\end{abstract}



\maketitle 

\section{Introduction}
\label{sect:Intro}

The analysis of the data collected recently at RHIC suggests that matter created in ultra-relativistic heavy-ion collisions behaves like a perfect fluid \cite{Heinz:2005zg}. This hypothesis inspires new interests in the studies of relativistic hydrodynamics of perfect and viscous fluids \cite{Teaney:2003kp,Hirano:2005wx,Hirano:2005dc,Hirano:2005xf,Hama:2005dz,Eskola:2005ue,Heinz:2005bw,Nonaka:2005aj,Huovinen:2006jp,Andrade:2006yh,Baier:2006sr,Baier:2006um,Koide:2006ef,Nonaka:2006yn,Satarov:2006jq,Hirano:2007xd}. The aim of this paper is to study the effects of the temperature dependence of the sound velocity on the hydrodynamic evolution of matter created in ultra-relativistic heavy-ion collisions. We use the approach to the relativistic hydrodynamics of perfect fluid where the sound velocity is the only thermodynamic parameter entering the formalism \cite{Baym:1983sr,Chojnacki:2004ec,Chojnacki:2006tv}. In this way, we are able to observe and analyze the immediate consequences of a particular choice of the function $c_s(T)$ on the space-time evolution of matter. 

Our study may be regarded as complementary to Ref. \cite{Huovinen:2005}, where the effects of a particular choice of the equation of state on the particle spectra and elliptic flow were studied. Our approach differs from Ref. \cite{Huovinen:2005} by the selection of the equations of state under investigations. In our case the equations of state are always constrained by the lattice results in the high temperature limit and differ from each other mainly in the neighborhood of the critical temperature. On the other hand, the main differences between the equations of state studied in Ref. \cite{Huovinen:2005} appear in the region above the critical temperature. 

In our approach the full information about the equation of state is contained in the temperature dependent sound velocity. At high temperatures,  $T > 1.15 \,T_c$, we use the sound velocity function obtained from the recent lattice simulations of QCD \cite{Aoki:2005vt}, whereas at low temperatures, $T < 0.15\,T_c$,  we use the result of the ideal hadron gas model \cite{Chojnacki:2004ec}. At moderate temperatures different interpolations between those two results are employed. The interpolating functions have a local maximum at $T = 0.4 \, T_c$ (corresponding to the maximal value of the sound velocity in the hadron gas) and a local minimum at $T=T_c$ (corresponding to the expected minimal value of the sound velocity at the phase transition). The values of the sound velocity at the maximum and minimum are chosen in such a way that in the limit of very high temperatures all considered sound-velocity functions yield the entropy density consistent with the lattice simulations of QCD. 

We emphasize that in all the considered cases we take into account only smoothly varying functions $c_s(T)$, hence we deal with the cross-over phase transitions rather than with the rigorous phase transitions of a given order that might be determined by the discontinuity of the appropriate thermodynamic variable. Nevertheless, the deeper is the minimum of the sound velocity at the critical temperature, the more the system evolution resembles the behavior typical for the exact first order phase transition. The main difference between the first order phase transition studied commonly in the literature and the {\it approximate} first-order phase transition studied below (as one of the examples) is that in our case the phase transition does not lead to the ideal quark-gluon plasma but rather to strongly interacting plasma characterized by the lattice results.

\begin{figure}[t]
\begin{center}
\includegraphics[angle=0,width=0.5\textwidth]{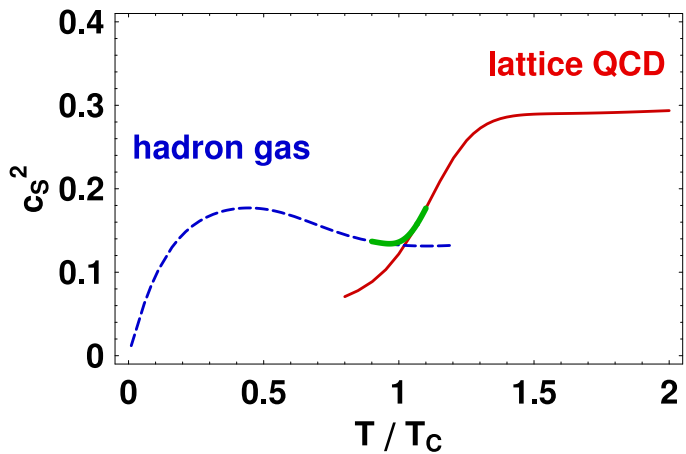}
\end{center}
\caption{Temperature dependence of the square of the sound velocity at zero baryon density. The plot shows the result of the lattice simulations of QCD \cite{Aoki:2005vt} (solid line) and the result obtained in the ideal hadron gas model \cite{Chojnacki:2004ec} (dashed line). A piece of the thick solid line describes the simplest interpolation between the two calculations. The critical temperature $T_c$ equals 170 MeV.}
\label{fig:cs2HGQCD}
\end{figure}
\begin{figure}[t]
\begin{center}
\includegraphics[angle=0,width=0.465\textwidth]{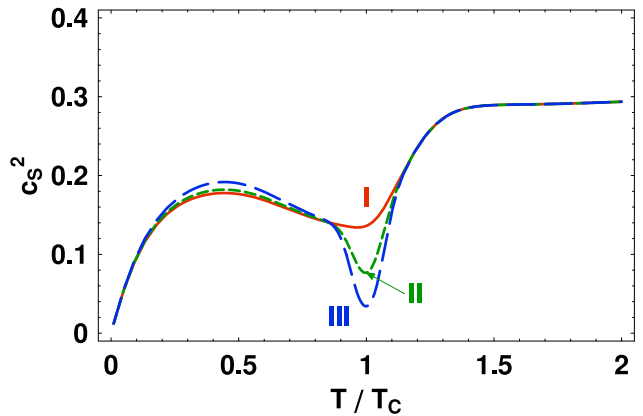}
\end{center}
\caption{Three different forms of the sound velocity used in the present hydrodynamic calculations. The solid line describes the interpolation between the lattice and the hadron-gas results \cite{Chojnacki:2004ec} with a shallow minimum where $c_s^2 = 0.14$  (case I), the long-dashed line describes the interpolation with a dip where $c_s^2 = 0.08$ (case II), finally the dashed line describes the interpolation with a deep minimum where $c_s^2 = 0.03$ (case III). In order to render the same entropy density at high temperatures, in the cases II and III the sound velocity in the region $T \approx 0.4 \,T_c$ is slightly increased compared to that considered in the case I. The larger values of the sound velocity in this region may be attributed to the repulsive van der Waals effects. }
\label{fig:cs2deep}
\end{figure}

In spite of such differences, in agreement with earlier studies \cite{Rischke:1995ir,Hung:1995eq,Rischke:1995cm,Rischke:1996em} we find that a deep minimum of the sound velocity leads to the very long, exceeding 20 fm ($c=1$), evolution time of the system. We further find that even moderate minima lead to quite long evolution times of the system, which exceed 15 fm.  Clearly, such long evolution times are not compatible with the recent HBT estimates of the lifetime of the system. For example, one finds times of about 10 fm using the RHIC data in the relation $R_L(m_T) = \tau \sqrt{T_k/m_T}$ \cite{Makhlin:1987gm}, which connects the longitudinal pion correlation radius $R_L$, the kinetic freeze-out temperature $T_k$, the evolution time $\tau$, and the transverse mass of the pion pair $m_T$. Another example is Ref. \cite{Csanad:2004mm} that obtained only 6 fm for the effective duration of the hydrodynamic evolution in Au+Au collisions at RHIC. 

Only if the sound velocity has a shallow minimum resulting from the simplest interpolation between the ideal hadron gas result at $T=0.85\,T_c$ and the lattice result at $T = 1.15\,T_c$, see Fig. \ref{fig:cs2HGQCD}, one finds that the evolution time of the system might be compatible with the HBT result. We thus conclude that the hydrodynamic description becomes adequate if the QCD cross-over phase transition renders the smooth temperature variations of the sound velocity, with a possible shallow minimum at $T_c$ where the values of $c_s^2(T)$ remain well above 0.1.

\begin{figure*}[t]
\begin{center}
\includegraphics[angle=0,width=0.75\textwidth]{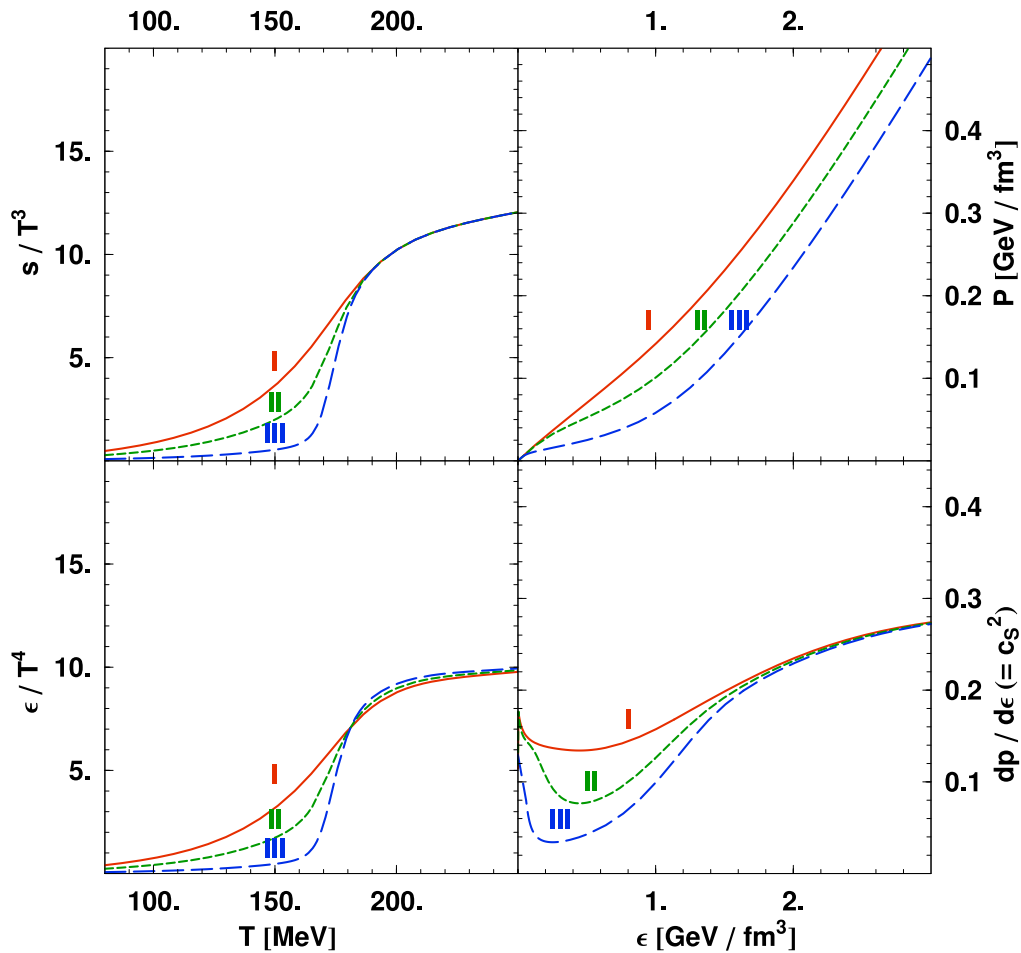}
\end{center}
\caption{The temperature dependence of the entropy density and energy density, panels a) and b), as well as the energy density dependence of the pressure and sound velocity, panels c) and d). One can observe that the deeper is the minimum of the sound velocity function, the steeper is the increase of the entropy density and the energy density.}
\label{fig:huo}
\end{figure*}

\section{Temperature dependent sound velocity}

In Fig. \ref{fig:cs2HGQCD} we show the function $c_s^2(T)$ obtained from the ideal hadron gas calculation \cite{Chojnacki:2004ec} and the lattice simulations of QCD \cite{Aoki:2005vt}. The ideal hadron gas result (dashed line) was obtained with the full mass spectrum of resonances \cite{Broniowski:2000bj,Broniowski:2004yh}. One can check that at very small temperatures the sound velocity is determined basically by the properties of the massive pion gas; for $T << m_\pi$, where $m_\pi$ is the pion mass, one finds $c_s(T) = \sqrt{T/m_\pi}$. Clearly, this behavior is quite different from the limiting value $c_s = 1/\sqrt{3}$ characterizing the massless pion gas. The lattice calculations (solid line) were obtained for physical masses of the light quarks and the strange quark \cite{Aoki:2005vt}. One can see that the results of the two calculations cross around the critical temperature, we assume $T_c$ = 170 MeV, and may be naturally interpolated (see a thick solid line in this region). Taking the lattice result at the face value, one expects that the sound velocity significantly drops down in the region $T \approx T_c$. Similar behavior, with $c_s(T_c)$ reaching zero,  is expected in the case of the first order first transition where the changes of the energy happen at constant pressure. However, the lattice simulations suggest that for three massive quarks with realistic masses we deal with the cross-over rather than with the first order phase transition, hence the sound velocity remains finite, as is consistently shown in Fig.  \ref{fig:cs2HGQCD}. Nevertheless, the exact values of the sound velocity in the region $T \approx T_c$ are in our opinion poorly known, since the lattice calculations are not very much reliable for $T < T_c$ and, at the same time,  the use of the hadron gas model with vacuum parameters becomes unrealistic for large densities (temperatures) \footnote{The authors of Ref. \cite{Aoki:2005vt} state that in the hadronic phase the lattice spacing is larger than 0.3 fm and the lattice artifacts cannot be controlled in this region}. In this situation, it is practical to consider different interpolations between the lattice and hadron-gas results and to analyze the physical effects of a particular choice of the interpolating function. 

In this paper we consider three different sound-velocity functions $c_s(T)$. Below, we refer to these three options as to the cases I, II and III, see Fig. \ref{fig:cs2deep}. In the case I,  we use the sound-velocity function which agrees with the ideal hadron gas model of Ref. \cite{Chojnacki:2004ec} in the temperature range $0 < T < 0.85 \,T_c$ and with the lattice result in the temperature range  $T > 1.15 \,T_c$. In the region close to the critical temperature, $0.85 \,T_c < T < 1.15 \,T_c$, a simple interpolation between the two results is used. We have checked that such a simple interpolation yields directly the entropy density consistent with the lattice result. Namely, the use of the thermodynamic relation 
\begin{equation}
s(T) = s(T_0) \exp\left[ \,\,\,\int\limits_{T_0}^T \,\frac{dT^\prime}{T^\prime c_s^2(T^\prime)}
\right],
\label{sumrule}
\end{equation}
relating the entropy density with the sound velocity for zero baryon chemical potential, gives the function $s(T)$ which agrees with the lattice result at high temperatures, $s(T)/T^3 \approx 12$ at $T = 1.5 \, T_c$ \cite{Aoki:2005vt}. 

In the cases II and III, the sound-velocity interpolating functions have a distinct minimum at \mbox{$T = T_c$}. Comparing to the case I [with \mbox{$c_s(T_c)=0.37$} and \mbox{$c_s^2(T_c)=0.14$}], the value of the sound velocity at \mbox{$T = T_c$} is reduced by 
\mbox{25 \%} in the case II [where \mbox{$c_s(T_c)=0.28$} and \mbox{$c_s^2(T_c)=0.08$}], and by 50\% in the case III [where \mbox{$c_s(T_c)=0.19$} and \mbox{$c_s^2(T_c)=0.03$}].  From Eq. (\ref{sumrule}) one concludes that the relative decrease of the sound velocity at $T_c$ leads to the relative increase of entropy density for high temperatures. Hence, in order to have the same value of the entropy density at high temperatures, a decrease of the sound velocity function in the region $T \approx  T_c$ should be compensated by its increase in a different temperature range. For our interpolating functions in the cases II and III we assume that the values of $c_s(T)$ in the range $0.15 \,T_c < T < 0.85 \,T_c$ are slightly higher than in the case I, see Fig. \ref{fig:cs2deep}. Such modifications may be regarded as the parameterization of the repulsive van der Waals forces in the hadron gas. The values of the maxima are chosen in such a way that the entropy densities for three considered cases are consistent with the lattice result, see the upper left panel of Fig. \ref{fig:huo} where the functions $s(T)/T^3$ are shown. 

We stress that in the three considered cases the values of $c_s(T)$ in the temperature range $T_c < T < 1.25 \, T_c$ remain significantly below the massless limit $1/\sqrt{3}$. Such a limiting value is implicitly  used in many hydrodynamic codes assuming the equation of state of non-interacting massless quarks and gluons for $T > T_c$, see for example the extended 3+1 hydrodynamic model of Ref. \cite{Nonaka:2006yn}. We expect that the cooling of the central part of our system will be significantly slower than the cooling of the systems containing an ideal quark-gluon plasma in the core and this effect reflects the non-perturbative phenomena taking place above $T_c$ that may be attributed to the formation of strongly interacting quark-gluon plasma. We also note that our study differs from the 3+1 model of Ref. \cite{Hama:2005dz} where the interpolation between the equation of state of the ideal quark-gluon plasma above $T_c$ (not of the lattice QCD) and the resonance gas below $T_c$ was introduced. We have checked that the corrections introduced in this way differ from the non-perturbative effects found in the lattice simulations. Moreover, Fig. 3 of Ref. \cite{Hama:2005dz} (lower left panel) indicates that the sound velocity at low temperatures does not drop to zero as in our case.

\section{Hydrodynamic equations}

In this Section we recapitulate the main features of approach to the relativistic hydrodynamics, which is the generalization of the method introduced by Baym et al. \cite{Baym:1983sr} and has been developed in Ref. \cite{Chojnacki:2004ec} (introduction of the temperature dependent sound velocity and initial transverse flow) and Ref. \cite{Chojnacki:2006tv} (inclusion of the cylindrical asymmetry). We rewrite the equations in such a way that the sound velocity is the only thermodynamic parameter characterizing the matter. Such reformulation of the hydrodynamic equations is always possible in the case of zero net baryon density. Since we consider the evolution of matter formed in the central region at the largest RHIC energies, the approximation of zero net baryon density is reasonable. Thermal analysis of the ratios of hadron multiplicities indicates that the baryon chemical potential at RHIC top energies is about 30 MeV, i.e., it is much smaller than the corresponding temperature of about 170 MeV \cite{Florkowski:2001fp,Torrieri:2004zz}. We also restrict our considerations to the boost-invariant and cylindrically symmetric expansion. The assumption of boost-invariance is again good for the central region, while the effects of azimuthal asymmetry are typically small, of the order of 10\%, and have no effects on our conclusions. 

In the case of vanishing baryon chemical potential the hydrodynamic equations may be written in the following form
\begin{equation}
u^{\mu }\partial _{\mu }\left( T\,u^{\nu }\right) =\partial ^{\nu }T,
\label{acc}
\end{equation}
\begin{equation}
\partial _{\mu }\left(s u^{\mu }\right) =0,
\label{ent}
\end{equation}
where $T$ is the temperature, $s$ is the entropy density, and $u^{\mu
}=\gamma \left( 1,\mathbf{v}\right) $ is the hydrodynamic four-velocity
(with $\gamma =1/\sqrt{1-v^{2}}$).  Since temperature is the only independent thermodynamic variable, all other thermodynamic quantities can be obtained from pressure given as a function of temperature, $P = P\left( T\right)$. Such a function plays a role of the equation of state. With the help of the thermodynamic relations
\begin{equation}
d\varepsilon = T ds ,\qquad dP = s dT,\qquad \varepsilon +P = T s,
\label{thermo}
\end{equation}
other thermodynamic quantities may be obtained. In addition, the equation of
state allows us to calculate the sound velocity
\begin{equation}
c_{s}^{2}=\frac{\partial P}{\partial \varepsilon }=\frac{s}{T}\frac{%
\partial T}{\partial s }.
\label{cs}
\end{equation}
From Eq. (\ref{cs}) one immediately finds Eq. (\ref{sumrule}) discussed in Sect. II.
The complete set of the thermodynamic quantities for the cases I, II and III is shown in Fig. \ref{fig:huo}.

For the boost-invariant systems with cylindrical symmetry Eq. (\ref{ent}) and the spatial components of Eq. (\ref{acc}) may be rewritten  as
\begin{eqnarray}
v_{r}\frac{\partial \ln T}{\partial t}+\frac{\partial \ln T}{\partial r}+%
\frac{\partial \alpha }{\partial t}+v_{r}\frac{\partial \alpha }{\partial r}
 & = &0, \label{accf}  \\
\frac{\partial \ln s }{\partial t}+v_{r}\frac{\partial \ln s }{%
\partial r}+v_{r}\frac{\partial \alpha }{\partial t}+\frac{\partial \alpha }{%
\partial r}+\frac{1}{t}+\frac{v_{r}}{r} & = &0,
\label{entf}
\end{eqnarray}
where $\alpha$ is the transverse rapidity of the fluid defined by the condition $v_{r}=\tanh\alpha$. The longitudinal component has the well known boost-invariant form $v_{z}=z/t$ \cite{Bjorken:1983qr}.
By introducing  the potential $\Phi \left( T\right) $ defined by the differentials
\begin{equation}
d\Phi =\frac{d\ln T}{c_{s}}=c_{s}d\ln s ,
\label{phi}
\end{equation}
and by the use of  the two functions $A_{\pm }$ defined by the formula
\begin{equation}
A_{\pm } = \Phi \pm \alpha,
\label{apm}
\end{equation}
Eqs. (\ref{accf}) and (\ref{entf}) become
\begin{eqnarray}
&&
\frac{\partial A_{\pm }\left( t,r\right) }{\partial t} +\frac{v_{r}\pm c_{s}}
{1\pm v_{r}\,c_{s}}\frac{\partial A_{\pm }\left( t,r\right) }{\partial r}\,
\nonumber \\
&&
+ \frac{c_{s}}{1\pm v_{r}\,c_{s}}\left( \frac{v_{r}}{r}+\frac{1}{t}\right) = 0.
\label{eqapm}
\end{eqnarray}
We note that Eq. (\ref{eqapm}) agrees with the formalism discussed in \cite{Baym:1983sr} if one defines $a_\pm = \exp(A_\pm)$. If the functions $A_{\pm}$ are known, the potential $\Phi$ may be calculated from the formula
\begin{equation}
\Phi = \frac{1}{2} \,\left( A_{+} + A_{-}\right),
\label{phioft}
\end{equation}
and the velocity is obtained from the equation
\begin{equation}
v_{r}  = \hbox{tanh} \,\left[ \frac{1}{2} \,\left( A_{+} - A_{-}\right)\right].
\label{vofa}
\end{equation}
The knowledge of the function $c_s(T)$ allows us, by the integration of 
Eq. (\ref{phi}), to determine $\Phi$ as a function of the temperature; 
this function will be called later $\Phi_T$. However, to get a closed system of 
equations for $A_+$ and $A_-$, we have to invert this relation and obtain $T$ as 
a function of $\Phi$; this function will be called later $T_\Phi$. In this way, 
the sound velocity may be expressed in terms of the functions $A_+$ and $A_-$,
\begin{equation}
c_s=c_s\left[T_\Phi\left( \frac{1}{2} \,\left( A_{+} + A_{-}\right) \right)\right],
\label{csapam}
\end{equation}
and Eqs. (\ref{eqapm}) may be solved numerically using the standard methods. 
The only restriction of our formalism is the condition against the formation of shock waves \cite{Baym:1983sr,Blaizot:1987cc}
\begin{equation}
 1 - c_s^2 + c_s T \frac{dc_s}{dT} 
= {d \over dT} \left({s \, c_s \over T } \right) \ge 0.
\label{stab1}
\end{equation}
We have checked that this condition is fulfilled by our sound-velocity profiles I, II and III.

For boost-invariant and cylindrically symmetric systems the entropy conservation law Eq. (\ref{ent}) implies that the following quantity is conserved in time
\begin{equation}
S = 2 \pi \int\limits_0^\infty dr \, r \, t \, \gamma(t,r) \, s(t,r) = \hbox{const}.
\label{globentropy}
\end{equation}
For one-dimensional expansion this condition is reduced to the famous Bjorken relation
$s(t) = s(t_0) t_0/t$. We have checked that the condition (\ref{globentropy}) is fulfilled in our calculations with very high accuracy, hence, no artificial entropy production is present in our algorithm. 

%%%%%%%%%%%%%%%%%%%%%%%%%%%%%%%%%%%%%%%%%%%%%%%%%%%%%%%%%%%%%%%%%%%%%%%%%%%%%%%%%%%%%
\section{Initial conditions}
%%%%%%%%%%%%%%%%%%%%%%%%%%%%%%%%%%%%%%%%%%%%%%%%%%%%%%%%%%%%%%%%%%%%%%%%%%%%%%%%%%%%%

For symmetry reasons, both the velocity field $v_r$ and the temperature gradient $\partial T/\partial r$ should vanish at $r=0$. This condition is achieved if the functions $A_+$ and $A_-$ are initially determined by a single function $A(r)$ according to the prescription  \cite{Baym:1983sr}
\begin{eqnarray}
A_+(t=t_0,r)=A(r), \quad A_-(t=t_0,r)=A(-r).
\label{inita}
\end{eqnarray}
Our main physical assumption about the initial state is that the initial temperature profile is connected with the nucleon-nucleus thickness function $T_A(r)$ by the following equation
\begin{equation}
T(t_0,r) =  T_s\left[ s(t_0,r) \right]   =
T_s\left[ s_0 \frac{T_A(r)}{T_A(0)} \right]   .
\label{T01}
\end{equation}
Here $T_s(s) $ is the inverse function to the entropy density function $s(T)$ and the parameter $s_0$ is the initial entropy at the center of the system. The idea to use Eq. (\ref{T01}) follows from the assumption that the initially produced entropy density  $s(t_0,r)$ is proportional to the density of wounded nucleons at a distance $r$ from the collision center \cite{Kolb:2003dz}. We use the value $s_0=70.5 \hbox{fm}^{-3}$ which yields \mbox{$T(t=t_0,r=0) = 2 T_c$}. 
We note that the functions $s(T)$ and $T_s(s)$ (evaluated for the cases I, II, and III) agree if the temperature or entropy is sufficiently large. Hence, in the three considered cases the 
initial temperature profiles are practically the same in the center of the system. The small differences appear however if we consider larger values of $r$ where the temperature and entropy drops down. We also note the global entropy defined by Eq. (\ref{globentropy}) is exactly the same in the cases I, II and III.

We recall that the thickness function is defined by the equation
\begin{equation}
T_A(r) = 2 \int dz \, \rho\left(\sqrt{r^2+z^2}\right),
\label{TA}
\end{equation}
where the function $\rho(r)$ is the nuclear density profile given by the Woods-Saxon function with a conventional choice of the parameters: $\rho_0 = 0.17 \,\hbox{fm} ^{-3}$,
\mbox{$r_0 = (1.12 A^{1/3} -0.86 A^{-1/3})$ fm}, $\, a = 0.54 \,\hbox{fm}, \, A = 197$. 
We note that the initial condition (\ref{T01}) may be included in the initial form of the function $A(r)$ 
\begin{equation}
A(t=t_0,r) = \Phi_T \left\{
T_s\left[ s_0 \frac{T_A(r)}{T_A(0)} \right] \right\} . 
\label{initaT}
\end{equation}

\begin{figure}[b]
\begin{center}
\includegraphics[angle=0,width=0.45\textwidth]{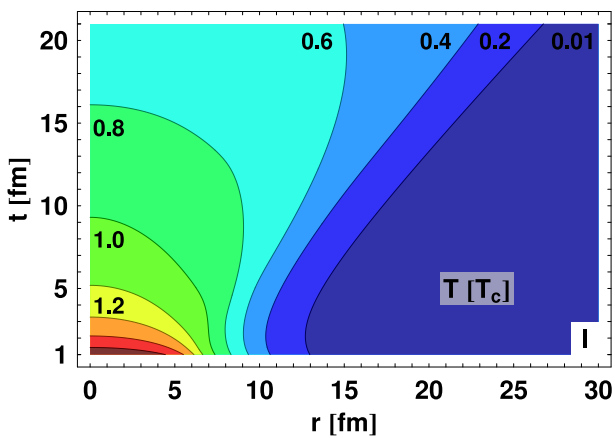}
\end{center}
\caption{Isotherms describing the hydrodynamic evolution for the case I; sound velocity with a shallow minimum where $c_s(T_c)$ = 0.37. The numbers denote the values of the temperature in the units of the critical temperature $T_c$ = 170 MeV. }
\label{fig:nodip}
\end{figure}

\section{Results}

\begin{figure}[t]
\begin{center}
\includegraphics[angle=0,width=0.45\textwidth]{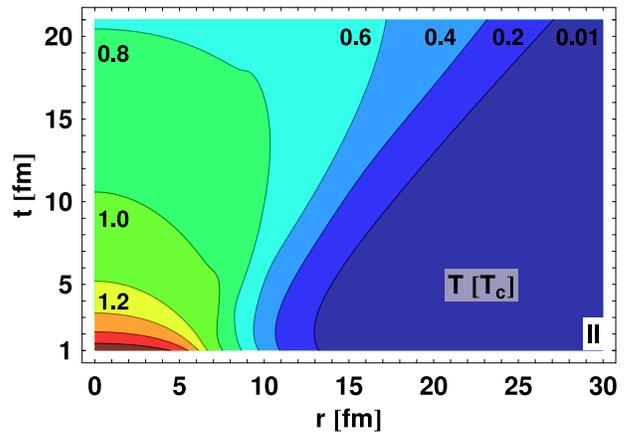}
\end{center}
\caption{Same as Fig. \ref{fig:nodip} but for the case II; sound velocity with a moderate minimum, $c_s(T_c)$ = 0.28.}
\label{fig:25dip}
\end{figure}

In Fig. \ref{fig:nodip} we show the isotherms describing the hydrodynamic evolution of the system with the sound velocity I. The numbers at the isotherms give the values of the temperature in the units of the critical temperature $T_c$. We observe that the center of the system cools down to 0.8 $T_c$ after the evolution time of about 15 fm. We note that the hydrodynamic description should be replaced (around $T \sim 0.8 \, T_c$) by the model describing hadronic rescattering (see, e.g., Ref. \cite{Nonaka:2005aj}) whose presence additionally increases the lifetime of the system. Another option is to assume that freeze-out happens at high temperature (see Refs. \cite{Broniowski:2001we,Broniowski:2002nf} where many physical observables were successfully reproduced under the assumption of a universal freeze-out taking place at the temperature of 165 MeV). In the latter case the lifetime of the system may be identified simply with the time when the system passes the phase transition. In the discussed case this time is of about 10 fm.

\begin{figure}[t]
\begin{center}
\includegraphics[angle=0,width=0.45\textwidth]{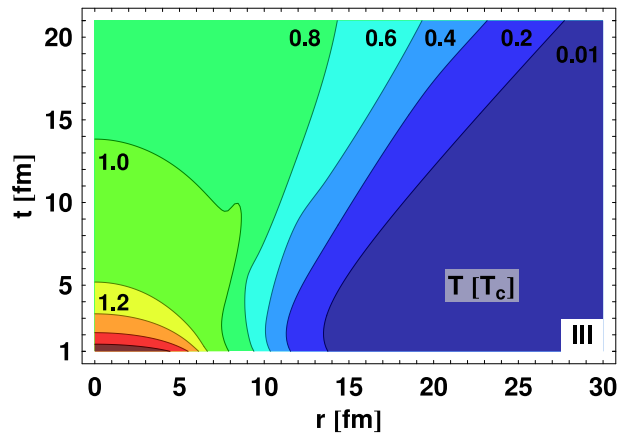}
\end{center}
\caption{Same as Fig. \ref{fig:nodip} but for the case III; sound velocity with a deep minimum, $c_s(T_c)$ = 0.19. }
\label{fig:50dip}
\end{figure}

In Fig. \ref{fig:25dip} we show the isotherms describing the hydrodynamic evolution of the system with the sound velocity II. The initial entropy is exactly the same as in the case I. One can notice that the central temperature does not drop below 0.8 $T_c$ before 20 fm. Clearly, the dip in the sound velocity causes a dramatic increase of the lifetime of the system. 

The most striking situation is presented in Fig. \ref{fig:50dip}. Here the sound velocity with the deepest minimum is considered. Again in this case the initial entropy of the system is exactly the same as in the cases I and II. One can notice that the system does not pass the phase transition before the considered evolution time of 20 fm. Contrary, even at that time the system has the tendency to expand more, the effect indicated by the shapes of the isotherms. The evolution time in the case III becomes longer than in the case II and of course much longer than in the case I. The situation depicted in Fig. \ref{fig:50dip} resembles very much the case of the first order phase transition, where the large latent heat is used to increase the volume of the system at constant temperature, see Fig. 3 panel b) where the step in the energy density as the function of the temperature is seen. Only when the latent heat is totally consumed, the center of the system starts further cooling. 

Interestingly, the case III describes how our approach with the temperature dependent sound velocity is capable of imitating the phenomena present at the real phase transitions. We have to remember, however, that the {\it approximate} first-order phase transition considered here differs from the real first-order phase transition between the hadron gas and the ideal quark-gluon plasma.

\section{Conclusions}

The observation of the effects related with the possible presence of the softest point of the QCD equation of state triggered the ideas about the long-living states formed in ultra-relativistic heavy-ion collisions \cite{Rischke:1995ir,Hung:1995eq,Rischke:1995cm,Rischke:1996em,Heinz:2005ja}. In view of the RHIC data indicating a short lifetime of the system we may analyze the problem of the lifetime in the reverse order, asking the question how soft the equation of the state may be to allow for a hydrodynamic description consistent with the HBT results. In our approach the issue of the softest point is translated  into the problem of realistic behavior of the sound velocity in the vicinity of the critical temperature. We argue that the smooth behavior of the sound velocity without any distinct minimum is favored if we demand the short hydrodynamic evolution time of about 10 fm. Our calculations show, as expected,  that the hydrodynamic evolution is very sensitive to the actual values of the sound velocity and indicates how extremely important is the good knowledge of the function $c_s(T)$. The reliable values of $c_s$ are not only required for the vicinity of the phase transition but also for smaller and larger temperatures, since the sound velocity plays a role of the coupling between the temperature gradients and the acceleration of the fluid. Finally, we note that the cooling of the system may be faster if the initial pre-equilibrium transverse flow is formed. The inclusion of such a flow in the initial conditions for the hydrodynamic equations leads naturally to faster expansion of matter.

\medskip

Acknowledgements: We thank Wojciech Broniowski for his very helpful comments and discussions.

%\bibliography{liter}

\begin{thebibliography}{10}
\expandafter\ifx\csname url\endcsname\relax
  \def\url#1{\texttt{#1}}\fi
\expandafter\ifx\csname urlprefix\endcsname\relax\def\urlprefix{URL }\fi

\bibitem{Heinz:2005zg}
U.~W. Heinz, nucl-th/0512051.

\bibitem{Teaney:2003kp}
D.~Teaney, Phys. Rev., {\bf C68} (2003) 034913, nucl-th/0301099.

\bibitem{Hirano:2005wx}
T.~Hirano, M.~Gyulassy, Nucl. Phys., {\bf A769} (2006) 71--94, nucl-th/0506049.

\bibitem{Hirano:2005dc}
T.~Hirano, Nucl. Phys., {\bf A774} (2006) 531--534, nucl-th/0511036.

\bibitem{Hirano:2005xf}
T.~Hirano, U.~W. Heinz, D.~Kharzeev, R.~Lacey, Y.~Nara, Phys. Lett., {\bf B636}
  (2006) 299--304, nucl-th/0511046.

\bibitem{Hama:2005dz}
Y.~Hama, {\it et~al.}, Nucl. Phys., {\bf A774} (2006) 169--178, hep-ph/0510096.

\bibitem{Eskola:2005ue}
K.~J. Eskola, H.~Honkanen, H.~Niemi, P.~V. Ruuskanen, S.~S. Rasanen, Phys.
  Rev., {\bf C72} (2005) 044904, hep-ph/0506049.

\bibitem{Heinz:2005bw}
U.~W. Heinz, H.~Song, A.~K. Chaudhuri, Phys. Rev., {\bf C73} (2006) 034904,
  nucl-th/0510014.

\bibitem{Nonaka:2005aj}
C.~Nonaka, S.~A. Bass, Nucl. Phys., {\bf A774} (2006) 873--876,
  nucl-th/0510038.

\bibitem{Andrade:2006yh}
R.~Andrade, F.~Grassi, Y.~Hama, T.~Kodama, J.~Socolowski, O., Phys. Rev. Lett.,
  {\bf 97} (2006) 202302, nucl-th/0608067.

\bibitem{Koide:2006ef}
T.~Koide, G.~S. Denicol, P.~Mota, T.~Kodama, hep-ph/0609117.

\bibitem{Nonaka:2006yn}
C.~Nonaka, S.~A. Bass, Phys. Rev., {\bf C75} (2007) 014902, nucl-th/0607018.

\bibitem{Hirano:2007xd}
T.~Hirano, U.~W. Heinz, D.~Kharzeev, R.~Lacey, Y.~Nara, nucl-th/0701075.


\bibitem{Huovinen:2006jp}
P.~Huovinen, P.~V. Ruuskanen, nucl-th/0605008.

\bibitem{Baier:2006sr}
R.~Baier, P.~Romatschke, U.~A. Wiedemann, Nucl. Phys., {\bf A782} (2007)
  313--318, nucl-th/0604006.

\bibitem{Satarov:2006jq}
L.~M. Satarov, I.~N. Mishustin, A.~V. Merdeev, H.~Stoecker, hep-ph/0611099.

\bibitem{Baier:2006um}
R.~Baier, P.~Romatschke, U.~A. Wiedemann, Phys. Rev., {\bf C73} (2006) 064903,
  hep-ph/0602249.



\bibitem{Baym:1983sr}
G.~Baym, B.~L. Friman, J.~P. Blaizot, M.~Soyeur, W.~Czyz, Nucl. Phys., {\bf
  A407} (1983) 541--570.

\bibitem{Chojnacki:2004ec}
M.~Chojnacki, W.~Florkowski, T.~Csorgo, Phys. Rev., {\bf C71} (2005) 044902,
  nucl-th/0410036.

\bibitem{Chojnacki:2006tv}
M.~Chojnacki, W.~Florkowski, Phys. Rev., {\bf C74} (2006) 034905,
  nucl-th/0603065.

\bibitem{Huovinen:2005}
P.~Huovinen, Nucl. Phys. {\bf A761} (2005) 296.


\bibitem{Aoki:2005vt}
Y.~Aoki, Z.~Fodor, S.~D. Katz, K.~K. Szabo, JHEP, {\bf 01} (2006) 089,
  hep-lat/0510084.

\bibitem{Rischke:1995ir}
D.~H. Rischke, S.~Bernard, J.~A. Maruhn, Nucl. Phys., {\bf A595} (1995)
  346--382, nucl-th/9504018.

\bibitem{Hung:1995eq}
C.~M. Hung, E.~V. Shuryak, Phys. Rev. Lett., {\bf 75} (1995) 4003--4006,
  hep-ph/9412360.

\bibitem{Rischke:1995cm}
D.~H. Rischke, M.~Gyulassy, Nucl. Phys., {\bf A597} (1996) 701--726,
  nucl-th/9509040.

\bibitem{Rischke:1996em}
D.~H. Rischke, M.~Gyulassy, Nucl. Phys., {\bf A608} (1996) 479--512,
  nucl-th/9606039.

\bibitem{Makhlin:1987gm}
A.~N. Makhlin, Y.~M. Sinyukov, Z. Phys., {\bf C39} (1988) 69.

\bibitem{Csanad:2004mm}
M.~Csanad, T.~Csorgo, B.~Lorstad, A.~Ster, J. Phys., {\bf G30} (2004)
  S1079--S1082, nucl-th/0403074.

\bibitem{Broniowski:2000bj}
W.~Broniowski, W.~Florkowski, Phys. Lett., {\bf B490} (2000) 223--227,
  hep-ph/0004104.

\bibitem{Broniowski:2004yh}
W.~Broniowski, W.~Florkowski, L.~Y. Glozman, Phys. Rev., {\bf D70} (2004)
  117503, hep-ph/0407290.

\bibitem{Florkowski:2001fp}
W.~Florkowski, W.~Broniowski, M.~Michalec, Acta Phys. Polon., {\bf B33} (2002)
  761--769, nucl-th/0106009.

\bibitem{Torrieri:2004zz}
G.~Torrieri, {\it et~al.}, Comput. Phys. Commun., {\bf 167} (2005) 229--251,
  nucl-th/0404083.

\bibitem{Bjorken:1983qr}
J.~D. Bjorken, Phys. Rev., {\bf D27} (1983) 140--151.

\bibitem{Blaizot:1987cc}
J.~P. Blaizot, J.-Y. Ollitrault, Phys. Lett., {\bf B191} (1987) 21--26.

\bibitem{Kolb:2003dz}
P.~F. Kolb, U.~Heinz, nucl-th/0305084.

\bibitem{Broniowski:2001we}
W.~Broniowski, W.~Florkowski, Phys. Rev. Lett., {\bf 87} (2001) 272302,
  nucl-th/0106050.

\bibitem{Broniowski:2002nf}
W.~Broniowski, A.~Baran, W.~Florkowski, Acta Phys. Polon., {\bf B33} (2002)
  4235--4258, hep-ph/0209286.

\bibitem{Heinz:2005ja}
U.~W. Heinz, J. Phys. Conf. Ser., {\bf 50} (2006) 230--237, nucl-th/0504011.

\end{thebibliography}

\end{document}